# The Prevalence of Misreporting and Misinterpreting Correlation Coefficients in Biomedical Literature

Jiayang Xu,[1] Xintong Chen,[1] Yufeng Liu,[1] Xiaoli Guo,[1] Shanbao Tong,[1]*

1. School of Biomedical Engineering, Shanghai Jiao Tong University, Shanghai, China.

**\* Corresponding author:**
Dr. Shanbao Tong, e-mail: stong@sjtu.edu.cn

**Abstract**
Correlation coefficient is widely used in biomedical and biological literature, yet its frequent misuse and misinterpretation undermine the credibility and reproducibility of the scientific findings. We systematically reviewed 1326 records of correlation analyses across 310 articles published in *Science*, *Nature*, and *Nature Neuroscience* in 2022. Our analysis revealed a troubling pattern of poor statistical reporting and inferring: 58.71% (95% CI: [53.23%, 64.19%], 182/310) of studies did not explicitly report sample sizes, and 98.06% (95% CI: [96.53%, 99.60%], 304/310) failed to provide confidence intervals for correlation coefficients. Among 177 articles inferring correlation strength, 45.25% (95% CI: [38.42%, 53.10%], 81/177) relied solely on point estimates, while 53.63% (95% CI: [46.90%, 61.58%], 96/177) drew conclusions based on null hypothesis significance testing. This widespread omission and misuse highlight a systematic gap in both statistic literacy and editorial standards. We advocate clear reporting guidelines mandating effect sizes and confidence intervals in correlation analyses to enhance the transparency, rigor, and reproducibility of quantitative life sciences research.

**Introduction**

Correlation coefficient, a statistical method for examining relationships between two variables, has shaped scientific inquiry since the late 19th century, when pioneers such as Francis Galton and Karl Pearson established the mathematical foundation for measuring associations (*1–3*). Today, correlation remains a cornerstone of statistical practice across the nature and social sciences (*4–7*). It enables researchers to identify and quantify associations, generate hypotheses, construct predictive models, and even explore potential causal mechanisms.

Yet, equating correlation with causation is a well-recognized statistical fallacy, as it risks ignoring confounding variables, oversimplifying multidimensional interactions, and leading to erroneous conclusions. In biomedical research, integrating statistical inference into correlation analysis presents additional challenges. A central limitation is the reliance on finite sample datasets, which inevitably fail to capture the full variability of the target population. As a result, population parameters must be approximated through point estimates, a process unavoidably affected by sampling variability. In this context, correlation analysis typically extrapolates population-level relationships from limited observations using inferential frameworks such as null hypothesis significance testing (NHST), confidence intervals, or Bayesian posterior probabilities.

Within the frequentist framework, confidence intervals provide a way to construct ranges that would capture the true parameter value in a high proportion of repeated samples. For correlation coefficients, confidence intervals therefore quantify the uncertainty involved in inferring the true association between two variables from sample data (*8, 9*). NHST, by contrast, assesses whether a sample correlation differs significantly from zero. Although both are widely used, neither provides the probability that a conclusion is correct. Their validity relies on the conceptual repetition of sampling and is governed by Type I and Type II error rates—assumptions that have long been the subject of debate in statistical methodology (*10, 11*). In contrast, Bayesian statistics directly quantifies the posterior probability of population parameters by combining prior information with the likelihood of the observed data (*12*). Despite offering a more coherent probabilistic framework and more intuitive interpretation than frequentist methods, Bayesian analysis remain less common in biomedical research—largely due to computational demand and entrenched conventions.

Even among top-tier journals, systematic reporting of confidence intervals has yet to be firmly established, and many conclusions continue to rely solely on point estimates from samples. Moreover, when confidence intervals or significance levels (i.e., *p*-values) for correlation coefficients are reported, misinterpretations are pervasive—for example, treating overlapping confidence intervals as evidence of no difference, or using a *p*-value as an indicator of correlation strength. For instance, a *Nature Neuroscience* paper reported high correlation between speech feature and clinical scores in amyotrophic lateral sclerosis (ALS) as below(*13*). All interpretations of Pearson's *r* and their comparisons were based solely on point estimates.

> *"Features derived from the voice tasks (single-breath count, read-aloud passage and free speech) each correlated highly with the bulbar subdomain of the ALS Functional Rating Scale-Revised (ALSFRS-R; Pearson's R = 0.8, slope = 1.14; Pearson's R = 0.89, slope = 0.98; and Pearson's R = 0.71, slope = 1.12, respectively).*
>
> *Importantly, semantic analysis of the picture description task was highly correlated with the ALS-Cognitive Behavioral Screen (CBS) (R = 0.72) and less correlated with*

Likewise, a *Science* article reported a "*strong linear relationship*" between paired-pulse short-term plasticity (STP) and cortical depth based solely on the *p*-value (*14*). Such reporting obscures the practical magnitude and uncertainty of the observed association.

> *"The paired-pulse STP showed a strong linear relationship with depth in the human data ($P = 5.4 \times 10^{-4}$), varying from weak facilitation for the most superficial cells ($0.02 \pm 0.03$, mean ± SEM) to depression for the deepest ($-0.32 \pm 0.09$).*
>
> *A strong correlation was also found with the AP upstroke/downstroke ratio of the presynaptic cell only ($P = 1.2 \times 10^{-4}$ versus $P > 0.3$ for postsynaptic)."*

As statisticians have cautioned, overreliance on NHST can cause to overlook the practical importance of the findings and to misinterpret nonsignificant results as "no effect" (*15*).

Such misuse and misinterpretation of correlation coefficients inevitably undermine the validity and reproducibility of research findings. These observations raise two questions:

*(1) How prevalent are such reporting and interpretive issues in leading scientific journals?*
*(2) What statistical principles should be emphasized to improve transparency and reproducibility?*

To address these questions, we conducted a retrospective review of all biomedical articles published in *Science*, *Nature*, and *Nature Neuroscience* during 2022. Our analysis focused on studies that employed Pearson's *r*, Spearman's *ρ*, or Kendall's *τ* for inferences and interpretation. Each article was evaluated along two dimensions: (1) the completeness of statistical reporting (including correlation coefficients, *p*-values, and confidence intervals) and (2) the basis of inferential interpretation. We identified two common forms of incorrect inference: (a) point-estimate-only interpretations that ignore estimation uncertainty, and (b) NHST-based inference that conflates statistical significance with effect size. Building on these findings, we propose practical guidelines for reporting, comparing, and interpreting correlation coefficients to promote more rigorous, transparent, and reproducible research across the life sciences.

## Results
### Overview
In total, we examined 1,760 research articles published in *Science*, *Nature*, and *Nature Neuroscience* in 2022. Of these, 1,095 fell within the domains of biology and biomedical engineering. Following a rigorous multi-stage screening, 310 articles were identified as employing at one form the correlation analysis, yielding 1,327 distinct instances for detailed evaluation (**Fig. 1**). Specifically, *Science* contributed 75 articles (272 cases), *Nature* 171 articles (657 cases), and *Nature Neuroscience* 64 articles (398 cases). Comprehensive details for each article and individual case are provided in supplementary material data S1-S3. These data formed the basis for our systematic assessment of reporting completeness and inferential accuracy in the use of correlation analysis.

### Reporting the key statistical parameters
Among the 310 reviewed articles, 84.51% (95% CI: [80.49%, 88.54%], 262/310) reported correlation coefficients (**Fig. 2A**), while 66.13% (95% CI: [60.86%, 71.40%], 205/310) included corresponding *p*-values (**Fig. 2B**). In contrast, only 41.29% (95% CI: [35.81%, 46.77%], 128/310) explicitly reported sample sizes (**Fig. 2C**), and a mere 1.94% (95% CI: [0.24%, 3.63%], 6/310) provided confidence intervals for correlation coefficients (**Fig. 2D**). The omission of sample sizes and confidence intervals severely constrains the interpretability and the reproducibility of these results.

### Accuracy of inferential interpretation
More than half of the examined articles (57.10%, 95% CI: [51.59%, 62.61%], 177/310) included explicit verbal interpretations of correlation strength, often using terms such as "strongly," "weakly," "moderately," "highly," or "lowly correlated." However, the underlying inferential logic was frequently flawed. Among these 177 papers, 45.25% (95% CI: [38.42%, 53.10%], 81/177) drew conclusions solely from point estimates of correlation coefficient without acknowledging estimation uncertainty (**Fig. 3A**). Another, 53.63% (95% CI: [46.90%, 61.58%], 96/177) relied exclusively on NHST, specifically the *p*-value, to infer strength (**Fig. 3B**), conflating statistical significance with the magnitude or importance of the observed relationship.

## Discussion
Correlation analysis remains one of the most frequently used statistical tools in biomedical research for examining associations between variables. Our comprehensive review indicates that roughly one in five published articles in biomedicine employs correlation analysis, reflecting its central role in data interpretation across the life sciences. However, the accuracy and credibility of conclusions derived from such analyses depend critically on their appropriate application and interpretation. Despite its ubiquity, our findings reveal widespread methodological and reporting shortcomings—even in leading journals—which compromise the reproducibility and reliability of published results. These observations underscore an urgent need to improve statistical literacy, editorial standards, and reporting practices related to correlation analysis. In this section, we discuss the underlying causes of these issues and propose practical recommendations for enhancing the rigor, transparency, and interpretability of correlation-based studies.

To ensure valid application, several prerequisites must be firstly considered before performing correlation analysis:

1. **Independence of samples.** Each observation in a dataset should represent an independent measurement. Repeated measurements from the same subject session, or experimental unit must not be treated as independent observations. Violations of this assumption can artificially inflate the apparent correlation strength and lead to spurious inferences. When data exhibit nested or hierarchical structure, such as repeated measures, multi-site studies, or family-based sampling, more appropriate modeling frameworks such as linear mixed-effects models should be applied (*16, 17*).

2. **Appropriate choice of method.** The selection of correlation coefficient must align with data type and distribution. Pearson's correlation is suitable for continuous variables that approximate a normal distribution and exhibit linear relationships (e.g., height, weight). Spearman's rank correlation, by contrast, is preferable for monotonic but nonlinear relationships or ordinal data. Kendall's $\tau$ offers a robust alternative for assessing concordance in ranked data and is particularly valuable in consistency or agreement studies where tied ranks frequently occur.

3. **Assumptions of data distribution.** While the computation of Pearson's correlation coefficient itself does not make assumptions about the data distribution, the inference procedures that accompany it, such as the estimation of confidence intervals or the calculation of *p*-values, depend pre-assumptions of data. These include linearity, approximate bivariate normality, and homoscedasticity. When these assumptions are violated, researchers should either employ nonparametric methods or apply data transformation. Ignoring such assumptions can lead to incorrect inference, misestimated uncertainty, and reduced generalizability of results.

4. **Influence of outliers. Outliers can substantially distort correlation estimates by exerting** disproportionate influence on the linear relationship between variables. Identifying and managing these points is therefore critical to maintaining analytic validity. Detection methods include standardized residuals, leverage statistics, Cook's distance, and visual inspection by scatterplot or boxplots. Once identified, researchers should verify whether outliers reflect genuine biological variation, measurement error, or data processing artifacts.   Robust approaches, such as rank-based or trimmed correlations, can mitigate the effects of extreme observations. In all cases, transparency in reporting how outliers are handled strengthens the credibility and reproducibility of the correlation analyses.

We strongly recommend **reporting both effect sizes and measures of uncertainty.** In addition to the correlation coefficient, authors should clearly report the corresponding sample size, confidence interval, and an interpretation of the effect's practical significance within the study context. Confidence intervals provide a direct quantification of estimation uncertainty and facilitate meaningful comparisons across studies, whereas point estimates alone may present a misleading impression of precision. Overreliance on null-hypothesis significance testing without accompanying effect-size information obscures the magnitude and variability of observed relationships (*18*, *19*). For clarity and standardization, we recommend following APA guidelines when reporting correlation results (*20*). For instance, the correlation between weight and height in a sample of 50 participants could be reported as: $r$ (48) = .45, $p$ = .002, 95% CI [.18, .66]. Such comprehensive reporting practices enhance transparency, reproducibility, and the interpretability of empirical findings across disciplines.

There is no doubt that point estimates are limited because they lose information about uncertainty when generalized from the sample to the population. This limitation largely explains the pervasive use of null-hypothesis significance testing (NHST) in the literature—whether through reporting only *p*-values or both correlation coefficients and *p*-values—in an attempt to provide inferential meaning. However, NHST is inherently weak, and sometimes misleading, when used to gauge the strength of an effect. Regardless of its limitations within the frequentist framework, the *p*-value itself does not convey effect size (*21*). In contrast, confidence intervals provide crucial context by capturing both the magnitude and uncertainty of an observed relationship, thereby allowing a more nuanced interpretation of effect sizes (*8*, *18*, *22*). Although confidence intervals are sometimes misunderstood—they do not represent the probability that a parameter lies within a specific range but rather indicate the proportion of intervals that would contain the true parameter if the experiment were repeated many times—they remain a more informative and transparent approach to inference. For example, in Study 1 (n = 12), the correlation was $r$ = .70, 95% CI: [.21, .91], $p$ = .011, while in Study 2 (n = 200), the correlation was likewise $r$ = .70, 95% CI: [.62,.76], $p$ < .001, even though both may yield extremely significant *p*-values, we can only conclude a strong correlation in Study 2, but not in Study 1. Alarmingly, our review of three top-tier journals revealed a striking neglect of confidence-interval reporting, with fewer than 2% of studies including this essential information. This omission undermines the reproducibility of published findings, as correlation coefficients represent sample estimates rather than true population parameters. Without confidence intervals to quantify precision, effect sizes are easily overinterpreted, and conclusions may fail to generalize beyond the study sample. To address this issue, we propose a minimum reporting standard for correlation analyses: researchers should always present the sample size, the correlation coefficient, and its corresponding confidence interval, aka mandating the APA reporting style(*20*). This reporting practice aligns with contemporary statistical recommendations and enables readers to evaluate both the strength and precision of reported associations, thereby enhancing the transparency and reproducibility of quantitative research.

In addition to the issues discussed above regarding the reporting and inference of correlation coefficients, other types of statistical misinterpretation also persist. For example, in a *Nature* article (*23*), the authors conflated two distinct concepts of the confidence interval for a correlation coefficient and the confidence band around a regression line. This confusion is evident in their statement, "*The 95% confidence interval of the correlation coefficient is displayed*" (legends of Fig. 1), whereas the accompanying figure actually shows the confidence band for the regression line.

Another recurring problem involves the misuse of *p*-values when comparing correlations across groups. For instance, in a *Science* paper (*14*), the authors concluded that a correlation existed only for the action potential (AP) upstroke/downstroke ratio of presynaptic cells, but not for postsynaptic cells, based solely on *p*-value comparisons. Their statement — "A strong correlation was also found with the AP upstroke/downstroke ratio of the presynaptic cell only ($P = 1.2 \times 10^{-4}$ versus $P > 0.3$ for postsynaptic)"—reveals a fundamental misunderstanding of null-hypothesis testing. *P*-values reflect the compatibility of observed data with the null hypothesis; a non-significant result (e.g., $P > 0.3$) does not provide evidence for the absence of an effect (*15*). Therefore, directly comparing *p*-values between groups is mathematically invalid. In statistical inference, a difference in significance levels does not imply a statistically significant difference between the underlying effects. In the case of (*15*), a statistically sound way to test differences in correlation strength between presynaptic and postsynaptic groups is to use Fisher's *z*-transformed confidence intervals or bootstrapped difference intervals. These methods directly evaluate whether two correlations differ significantly while accounting for sampling variability, rather than relying on indirect or invalid comparisons of *p*-values.

Building on the issue of misinterpreted correlation comparisons, we also observed that many articles employ heatmaps to display correlations across large datasets. While heatmaps provide an intuitive visual summary of pairwise relationships, they often appear without the accompanying statistical details required for rigorous interpretation. To ensure that such visualizations are both informative and statistically meaningful, we recommend that authors supplement heatmaps with numerical information—specifically, confidence intervals for the reported correlation coefficients—either in the main text or in the supplementary materials. This integration would create a smoother link between visual exploration and quantitative inference, substantially improving the rigor, transparency, and interpretability of results.

Last but not least, the terms "significant" and "significantly" are pervasively used in reporting correlation results, appearing in 23.6% (95% CI: [18.8%, 28.3%], 73/310) of articles. Among these, only 10.9% (95% CI = [3.8%, 18.1%], 8/73) explicitly qualified the term as "statistically significant" or "statistically significantly." The term "significant" itself is inherently ambiguous because it conflates statistical significance with the magnitude of a real effect—a widespread misconception that often leads to misinterpretation. To avoid such semantic confusion, we strongly recommend clearly distinguishing between statistical significance and effect size. Specifically, the unqualified use of the term "significant" should be replaced by the more precise expression "statistically significant," and should be accompanied by a report of the practical magnitude of the effect (i.e., the effect size) when describing inferential results.

## Materials and Methods

### Article searching

The workflow for the literature review and case analyses is outlined in **Fig.1**. We queried the Web of Science database (https://www.webofscience.com) for 2022 publications in Science, Nature, and Nature Neuroscience, using "journal title" and "publication year" as search criteria. The results were further refined with MeSH terms related to "biology/biomedical engineering." Full texts meeting these criteria were downloaded and systematically screened for correlation-related keywords, including "correla-", "Pearson", "Spearman", and "Kendall".

### Case collection

Each occurrence of a correlation analysis in the main text—including instances where multiple analyses were presented within a single subfigure—was treated as an independent case. For every case, we systematically documented the statistical method (Pearson $r$, Spearman $\rho$, or Kendall's $\tau$), sample size, the reported coefficient value ($r$, $\rho$, or $\tau$), the associated $p$-value, and the corresponding figure or table citations.

### Case analysis and article-level statistics

We analyzed each case along two main dimensions and categorized articles accordingly:

**1. Completeness of statistics reporting:** We assessed whether authors provided key statistics like correlation coefficient ($r$, $\rho$, or $\tau$), $p$-value, sample size ($n$), and confidence interval (CI). Reporting completeness was summarized at the article level: an article was considered to have reported a statistic if it appeared in any of its analyses.

**2. Basis of inferential interpretation:** We examined whether and how authors described correlation direction (positive/negative), strength (strong/moderate/weak), and on what basis such conclusions were drawn—whether from $r$, $p$-value, both, or CI). If any case within an article used a specific inferential approach, the article was classified under that category. Articles could thus fall into multiple categories. From this classification, we identified two recurring forms of incorrect inference: (1) **point-estimation-only inference**, where authors relied solely on correlation coefficient of the sample, without acknowledging uncertainty; and (2) **NHST-based inference**, where conclusions were drawn primarily from $p$-value or its combination with correlation coefficients.

Proportions were computed at the article level for each journal and then aggregated across the three journals. Nighty-five percent confidence intervals were estimated using either the asymptotic normal approximation or the Wilson score interval, depending on sample size and the proportion's proximity to the boundaries (0 or 1).

**References**

1. F. Galton, Regression Towards Mediocrity in Hereditary Stature. *The Journal of the Anthropological Institute of Great Britain and Ireland* **15**, 246–263 (1886).

2. K. Pearson, VII. Note on regression and inheritance in the case of two parents. *Proc. R. Soc. Lond* **58**, 240–242.

3. F. Galton, The British Association: Section II, Anthropology: Opening address by Francis Galton, F.R.S., etc., President of the Anthropological Institute, President of the Section. *Nature* **32**, 507–510 (1885).

4. S. Senthilnathan, Usefulness of correlation analysis. *Political Methods: Quantitative Methods eJournal* (2019).

5. S. Yadav, Correlation analysis in biological studies. *Journal of the Practice of Cardiovascular Sciences* **4**, 116–121 (2018).

6. A. Agrawal, K. Verma, N. shivhare, A Study of Nature of Correlation. *Journal of Advanced Research in Applied Mathematics and Statistics* **8**, 1–5 (2023).

7. H. A. Miot, Correlation analysis in clinical and experimental studies. *J Vasc Bras* **17**, 275–279 (2018).

8. D. G. Altman, Why we need confidence intervals. *World J Surg* **29**, 554–556 (2005).

9. A. M. Attia, Why should researchers report the confidence interval in modern research. *Middle East Fertil Soc J* **10**, 78–81 (2005).

10. A. Banerjee, U. B. Chitnis, S. L. Jadhav, J. S. Bhawalkar, S. Chaudhury, Hypothesis testing, type I and type II errors. *Ind Psychiatry J* **18**, 127–131 (2009).

11. J. P. A. Ioannidis, Why most published research findings are false. *PLOS Medicine* **2**, null (2005).

12. H. Jackson, Y. Shou, N. A. B. M. Azad, J. W. Chua, R. L. Perez, X. Wang, M. E. A. de Kraker, Y. Mo, A comparison of frequentist and Bayesian approaches to the Personalised Randomised Controlled Trial (PRACTical)-design and analysis considerations. *BMC Med Res Methodol* **25**, 149 (2025).

13. E. G. Baxi, T. Thompson, J. Li, J. A. Kaye, R. G. Lim, J. Wu, D. Ramamoorthy, L. Lima, V. Vaibhav, A. Matlock, A. Frank, A. N. Coyne, B. Landin, L. Ornelas, E. Mosmiller, S. Thrower, S. M. Farr, L. Panther, E. Gomez, E. Galvez, D. Perez, I. Meepe, S. Lei, B. Mandefro, H. Trost, L. Pinedo, M. G. Banuelos, C. Liu, R.

Moran, V. Garcia, M. Workman, R. Ho, S. Wyman, J. Roggenbuck, M. B. Harms, J. Stocksdale, R. Miramontes, K. Wang, V. Venkatraman, R. Holewenski, N. Sundararaman, R. Pandey, D.-M. Manalo, A. Donde, N. Huynh, M. Adam, B. T. Wassie, E. Vertudes, N. Amirani, K. Raja, R. Thomas, L. Hayes, A. Lenail, A. Cerezo, S. Luppino, A. Farrar, L. Pothier, C. Prina, T. Morgan, A. Jamil, S. Heintzman, J. Jockel-Balsarotti, E. Karanja, J. Markway, M. McCallum, B. Joslin, D. Alibazoglu, S. Kolb, S. Ajroud-Driss, R. Baloh, D. Heitzman, T. Miller, J. D. Glass, N. L. Patel-Murray, H. Yu, E. Sinani, P. Vigneswaran, A. V. Sherman, O. Ahmad, P. Roy, J. C. Beavers, S. Zeiler, J. W. Krakauer, C. Agurto, G. Cecchi, M. Bellard, Y. Raghav, K. Sachs, T. Ehrenberger, E. Bruce, M. E. Cudkowicz, N. Maragakis, R. Norel, J. E. Van Eyk, S. Finkbeiner, J. Berry, D. Sareen, L. M. Thompson, E. Fraenkel, C. N. Svendsen, J. D. Rothstein, Answer ALS, a large-scale resource for sporadic and familial ALS combining clinical and multi-omics data from induced pluripotent cell lines. *Nature Neuroscience* **25**, 226–237 (2022).

14. L. Campagnola, S. C. Seeman, T. Chartrand, L. Kim, A. Hoggarth, C. Gamlin, S. Ito, J. Trinh, P. Davoudian, C. Radaelli, M.-H. Kim, T. Hage, T. Braun, L. Alfiler, J. Andrade, P. Bohn, R. Dalley, A. Henry, S. Kebede, A. Mukora, D. Sandman, G. Williams, R. Larsen, C. Teeter, T. L. Daigle, K. Berry, N. Dotson, R. Enstrom, M. Gorham, M. Hupp, S. D. Lee, K. Ngo, P. R. Nicovich, L. Potekhina, S. Ransford, A. Gary, J. Goldy, D. McMillen, T. Pham, M. Tieu, L. Siverts, M. Walker, C. Farrell, M. Schroedter, C. Slaughterbeck, C. Cobb, R. Ellenbogen, R. P. Gwinn, C. D. Keene, A. L. Ko, J. G. Ojemann, D. L. Silbergeld, D. Carey, T. Casper, K. Crichton, M. Clark, N. Dee, L. Ellingwood, J. Gloe, M. Kroll, J. Sulc, H. Tung, K. Wadhwani, K. Brouner, T. Egdorf, M. Maxwell, M. McGraw, C. A. Pom, A. Ruiz, J. Bomben, D. Feng, N. Hejazinia, S. Shi, A. Szafer, W. Wakeman, J. Phillips, A. Bernard, L. Esposito, F. D. D'Orazi, S. Sunkin, K. Smith, B. Tasic, A. Arkhipov, S. Sorensen, E. Lein, C. Koch, G. Murphy, H. Zeng, T. Jarsky, Local connectivity and synaptic dynamics in mouse and human neocortex. *Science* **375**, eabj5861 (2022).

15. V. Amrhein, S. Greenland, B. McShane, Retire statistical significance. *Nature* **567**, 305–307 (2019).

16. E. Aarts, M. Verhage, J. V. Veenvliet, C. V. Dolan, S. Van Der Sluis, A solution to dependency: using multilevel analysis to accommodate nested data. *Nat Neurosci* **17**, 491–496 (2014).

17. S. E. Lazic, The problem of pseudoreplication in neuroscientific studies: is it affecting your analysis? *BMC Neurosci* **11**, 5 (2010).

18. G. Cumming, Understanding the new statistics: Effect sizes, confidence intervals, and meta-analysis. *Understanding the new statistics: Effect sizes, confidence intervals, and meta-analysis.*, xiv, 519–xiv, 519 (2012).

19. G. Cumming, The new statistics: Why and how. *Psychological Science* **25**, 7–29 (2014).

20. American Psychological Association, *Publication Manual of the American Psychological Association: The Official Guide to APA Style (7th Ed.)* (American Psychological Association).

21. R. Calin-Jageman, G. Cumming, From significance testing to estimation and Open Science: How esci can help. *International Journal of Psychology* **59**, 672–689 (2024).

22. A. Rovetta, L. Piretta, M. A. Mansournia, p-Values and confidence intervals as compatibility measures: guidelines for interpreting statistical studies in clinical research. *Lancet Reg Health Southeast Asia* **33**, 100534 (2025).

23. J.-B. Raina, B. S. Lambert, D. H. Parks, C. Rinke, N. Siboni, A. Bramucci, M. Ostrowski, B. Signal, A. Lutz, H. Mendis, F. Rubino, V. I. Fernandez, R. Stocker, P. Hugenholtz, G. W. Tyson, J. R. Seymour, Chemotaxis shapes the microscale organization of the ocean's microbiome. *Nature* **605**, 132–138 (2022).

**Acknowledgments**
We thank Jiaxing Li, Xintong Zhong, Limei Chen for their help in data collection.

**Declaration of Generative AI and AI assisted technologies in the writing process:** During the writing process of this work, the authors used ChatGPT in order to improve the language. After using the tool, the authors reviewed and edited the contents as needed and are solely responsible for the entire content, ideas, and conclusions of the paper.

**Funding:**   ST is partly supported by the MOST 2030 Brain Project (grant number: 2022ZD0208500) and National Natural Science Foundation of China (grant number: 82472097).

**Author contributions:**
Conceptualization: ST, XG
Methodology: JX, XC, ST, XG
Investigation: JX, XC, YL
Visualization: JX, XC
Supervision: ST, XG
Writing—original draft: JX, XC
Writing—review & editing: XG, ST

**Competing interests:** All other authors declare they have no competing interests.

# Figures and Tables

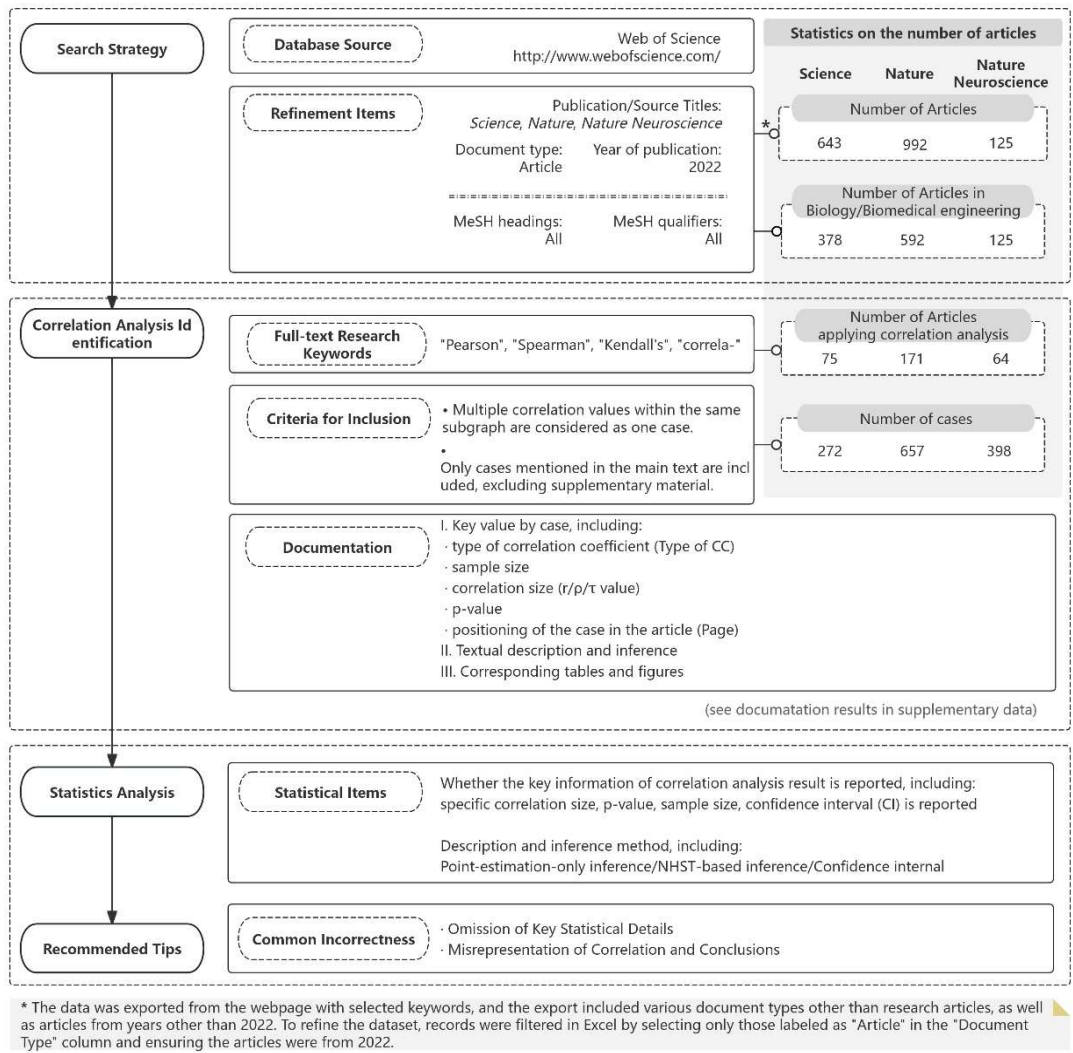

**Fig.1. Schematic of the research process.** The study began with a structured search of the Web of Science database using predefined criteria, including *publication type*, *year*, and *subject area*. Only records classified as "Article" were retained, while all other entry types from 2022 were excluded. The remaining publications were screened for correlation-related keywords (e.g., "Pearson," "Spearman," "Kendall's," "Corr"), and eligible studies underwent full-text review. The subsequent analysis assessed the frequency and quality of correlation reporting and identified inference incorrectness. Based on these findings, we formulated recommendations to improve statistical practice in correlation analyses. Finally, articles were categorized by subject area to illustrate their distribution and to summarize the number of studies retained at each stage.

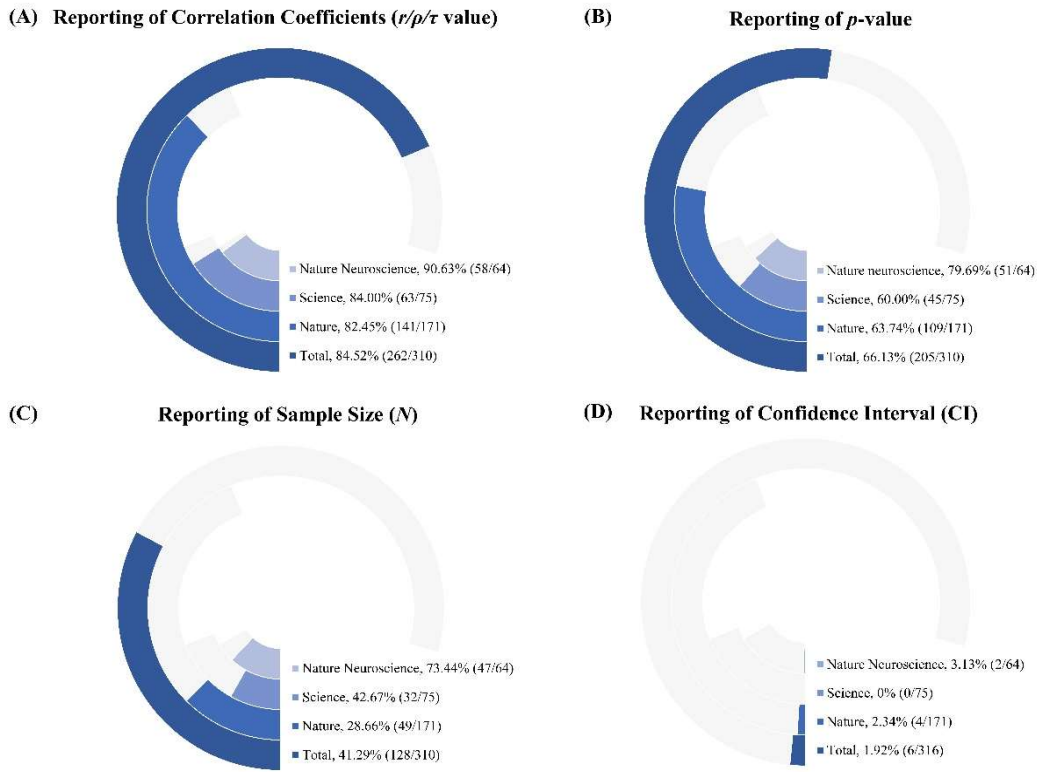

**Fig. 2. Reporting ratios of key statistical parameters in Science, Nature, and Nature Neuroscience.** (a) Correlation coefficients (r/ρ/τ), (b) p-values, (c) sample sizes (n), and (d) confidence intervals (CIs).

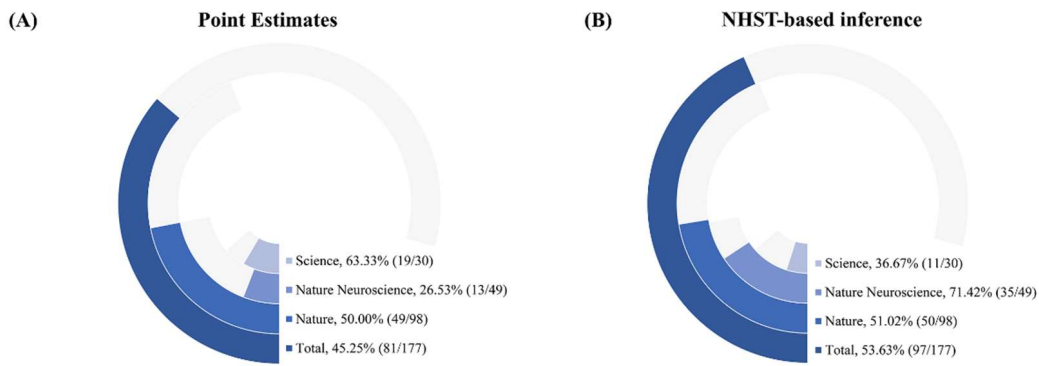

**Fig. 3. Reporting practices in correlation analyses across *Science*, *Nature*, and *Nature Neuroscience* in 2022.** The circular track plot illustrates how the strength of correlation was interpreted in the three journals. (A) point-estimation-only inference. (B) NHST-based inference.